\documentclass[hyperref,amsmath,amssymb,showpacs,floatfix,journal=ancac3,manuscript=article]{achemso}
\usepackage{graphicx}   
\usepackage{amsmath}    
\usepackage{amssymb}    
\usepackage{extarrows}
\usepackage{morefloats}
\usepackage[version=4]{mhchem} 
\usepackage{epstopdf} 
\DeclareGraphicsExtensions{.eps, .ps, .pdf}
\graphicspath{{ }{/home/ioan/floppy/corona/countries/.}{/home/ioan/floppy/corona/.}}

\newcommand{\gl}{equation} 
\newcommand{\gls}{equations}

\newcommand{\figname}{Fig.~} 
\newcommand{\figsname}{Figs.~}

\usepackage{mfirstuc}
\MFUnocap{of}
\MFUnocap{that}
\MFUnocap{by}
\MFUnocap{to}
\MFUnocap{on}
\MFUnocap{and}
\MFUnocap{in}
\MFUnocap{the}
\MFUnocap{for}
\MFUnocap{with}
\SectionNumbersOn
\author{Ioan B\^aldea}
\email{ioan.baldea@pci.uni-heidelberg.de}
\affiliation{Theoretical Chemystry, Heidelberg University, Im Neuenheimer Feld 229, D-69120 Heidelberg, Germany}

\title[Inaccuracies in Modeling COVID-19 Pandemic within a Continuous-Time Picture] 
      {
        Quantifying Inaccuracies in Modeling COVID-19 Pandemic 
        within a Continuous Time Picture}

\begin{document}
\label{firstpage}
\maketitle
\begin{abstract}
  Typically, mathematical simulation studies on COVID-19 pandemic forecasting
  are based on deterministic differential equations which assume that both 
  the number ($n$) of individuals in various epidemiological classes and the time ($t$)
  on which they depend are quantities that vary continuous. 
  This picture contrasts with the discrete representation of $n$ and $t$ underlying
  the real epidemiological data reported in terms daily numbers of infection cases,
  for which a description based on finite difference equations would be more adequate.
  Adopting a logistic growth framework, in this paper we present a quantitative analysis of the
  errors introduced by the continuous time description. This analysis
  reveals that, although the height of the epidemiological curve maximum is essentially
  unaffected, the position $T_{1/2}^{c}$ obtained within the continuous time representation
  is systematically shifted backwards in time with respect to
  the position $T_{1/2}^{d}$ predicted within the discrete time representation.
  Rather counterintuitively, the magnitude of this temporal shift
  $\tau \equiv T_{1/2}^{c} - T_{1/2}^{d} < 0$
  is basically insensitive to changes in infection rate $\kappa$. For a
  broad range of $\kappa$ values deduced from COVID-19 data at
  extreme situations (exponential growth in time and complete lockdown), we found 
  a rather robust estimate $\tau \simeq -2.65\,\mbox{day}^{-1}$.
  Being obtained without any particular assumption, the present mathematical
  results apply to logistic growth in general without any limitation to a specific
  real system. 
\end{abstract}
\section{Introduction}
\label{sec:intro}
Since the COVID-19 infectious disease caused by the
severe acute respiratory syndrome coronavirus 2 (SARS-CoV-2)
broke out in Wuhan City, the capital of Hubei province of China, \cite{WuhanOutbreak_WHO,WuhanOutbreak_Lancet}
on December 2019 and WHO declared it a pandemic \cite{WHO_COVID_19_Pandemic,WHO_COVID-19_Pandemic},
collaborative work was conducted worldwide by research institutions and pharma industry aiming at 
developing antiviral medication and vaccine in the global battle against this
unprecedented dangerous threat to planet's life. Along with such effort whose concrete results
are still to come, mathematical simulation can significantly assist governments in adopting the most adequate solutions
to suppress infection spread and minimize medical burden as well as dramatic economic and social consequences. 

Most population dynamics studies on COVID-19 pandemic as well as on
other epidemics carried out in the past
adopted a mathematical framework based on
a continuous time picture. This is particularly comfortable because it enables to derive the temporal
evolution of the populations of various epidemiological compartments by solving
a set ordinary differential equations. Still, a more realistic approach should aim at quantitatively
predicting or at least reproducing the new infection cases, and these are reported daily. Therefore,
a discrete time representation appears to be more adequate.
From this perspective, quantifying the errors arising from the 
simplified continuous time assumption is a significant question
that deserves consideration. However, to the best of author's knowledge, this issue
was not addressed in previous COVID-19 studies. For this reason,
it represents the focus of the present paper.

\section{The Logistic Model}
\label{sec:model}
Evolution in time of epidemics was extensively investigated in the literature, most often
within the celebrated
SIR model or many of its extensions \cite{Kermack:27,Kermack:32,Kermack:33,Bailey:75,Hethcote:94,Hethcote:00}.
An important problem with such approaches is the (too) large number of input parameters
they need in order to make reliable predictions, and their values are difficult to validate
before disease end \cite{IJIMAI-3841}.

For this reason, we will adopt here the simpler logistic growth model
\cite{Verhulst:1838,Verhulst:1847,Quetelet:1848,Ostwald:1883}.
Along with numerous applications in various other fields
\cite{McKendrick:1912,Lloyd:67,Cramer:02,Vandermeer:10,Baldea:2017m,Baldea:2018e},
this model turned out to be also useful in previous studies on epidemics
\cite{Mansfield:60,Waggoner:00,Koopman:04,Bangert:17},
also including COVID-19 pandemic \cite{Hermanowicz2020.03.31.20049486,Baldea:2020e,Baldea:2020f}.
As compared to various SIR-based approaches, the logistic growth model possesses a
great advantage: it can be directly validated against ongoing epidemiological reports
\cite{Baldea:2020f}.

In the continuous time description, logistic growth in time of an infected population
obeys a first-order differential equation
\begin{equation}
  \label{eq-dn-continuous}
\frac{d}{d t} n = \kappa n \left(1 - \frac{n}{N}\right)
\end{equation}
Here, $\kappa$ stands for the intrinsic infection rate, which characterizes the 
unlimited exponential (Malthus-type) population increase occurring in an
infinite environment. In real situations, exponential growth is suppressed
because the effective infection rate $\tilde{\kappa} = (d n/ dt)/n$ gradually decreases as the number of infections increases.
The parenthesis entering the RHS of \gl~(\ref{eq-dn-continuous})
resembles the Pauli blocking factor passionately discussed in charge transport in Fermi systems
\cite{Datta:92,Sols:92,Schoenhammer:93,Datta:97,Wagner:00}.
It represents the simplest possible way to
model a population dependent infection rate $\tilde{\kappa} = f(n)$, namely to assume that it linearly decreases as 
$n$ increases and eventually levels off at the maximum value $N$, which defines the so-called carrying capacity.

Epidemiologists refer to $n = n(t)$ as the total (cumulative) number of cases and to
$ \dot{n}(t) \equiv d n(t)/d t$ as the daily number of (new) infections.
The latter term stems from discrete time counterpart of the temporal derivative
\begin{equation}
  \label{eq-dn-discrete}
  \frac{d}{d t} n(t) 
  = \left. \frac{n(t + \Delta t) - n(t)}{\Delta t}\right\vert_{\Delta t \to 0}
  \approx \left. \frac{n(t + \Delta t) - n(t)}{\Delta t}\right\vert_{\Delta t = 1\,\mbox{(day)} }
  = n(t + 1) - n(t)
\end{equation}
By replacing the derivative entering the LHS of \gl~(\ref{eq-dn-continuous}) by its finite difference counterpart
of the RHS of \gl~(\ref{eq-dn-discrete}) we arrive at the discrete time formulation of logistic growth
\begin{equation}
  \label{eq-n-discrete}
  n(t + 1) - n(t) = \kappa\, n(t) \left[1 - \frac{n(t)}{N}\right]
\end{equation}
The advantage of the continuous time description of logistic growth
is that \gl~(\ref{eq-dn-continuous}) can be integrated out in closed analytical form
with the well known results
\begin{equation}
  \label{eq-n}
  n(t) = \frac{N}{1 + \exp\left[ - \kappa \left(t - T_{1/2}\right)\right]} 
\end{equation}

\begin{equation}
  \label{eq-dn}
  \dot{n}(t) = \frac{N\kappa}{4}\mbox{sech}^2\frac{\kappa \left(t - T_{1/2}\right)}{2}
\end{equation}
Notice that, whether in continuous or discrete time, in addition to the two
model parameters $\kappa$ and $N$,
an ``initial'' condition 
is needed to determine the time evolution $n = n(t)$ via \gls~(\ref{eq-dn-continuous})
or (\ref{eq-n-discrete}), respectively.
Most commonly this is imposed by
\begin{equation}
  \label{eq-n0}
n_0 = \left. n(t)\right\vert_{t=0}
\end{equation}
In continuous time description this is often recast
in terms of the half-time defined by $\left. n(t)\right\vert_{t=T_{1/2}} = N/2$
which yields
\begin{equation}
  \label{eq-half-time}
  T_{1/2} = \frac{1}{\kappa}\ln \left(\frac{N}{n_0} - 1\right)
\end{equation}

\section{Results and Discussion}
\label{sec:results}
Inspection of \gls~(\ref{eq-dn-continuous}) or (\ref{eq-n-discrete}) reveals that
results for the normalized number of cumulative cases  $n(t)/N$ and daily cases
($\dot{n}/N$ or $\left[n(t+1)-n(t)\right]/N$, respectively) as a function of the reduced time $t/T_{1/2}$
can be depicted as curves depending on a single parameter, namely $\kappa$; the value $n_0/N$ merely
serves to appropriately set the time origin.
To be specific, guided by our recent study \cite{Baldea:2020f},
in all the figures presented below we used $n_0/N \approx 1/1500$.
However, as just said, this particular choice does by no means affect the generality of the
results reported below.

\begin{figure}
  \centerline{
    \includegraphics[width=0.45\textwidth]{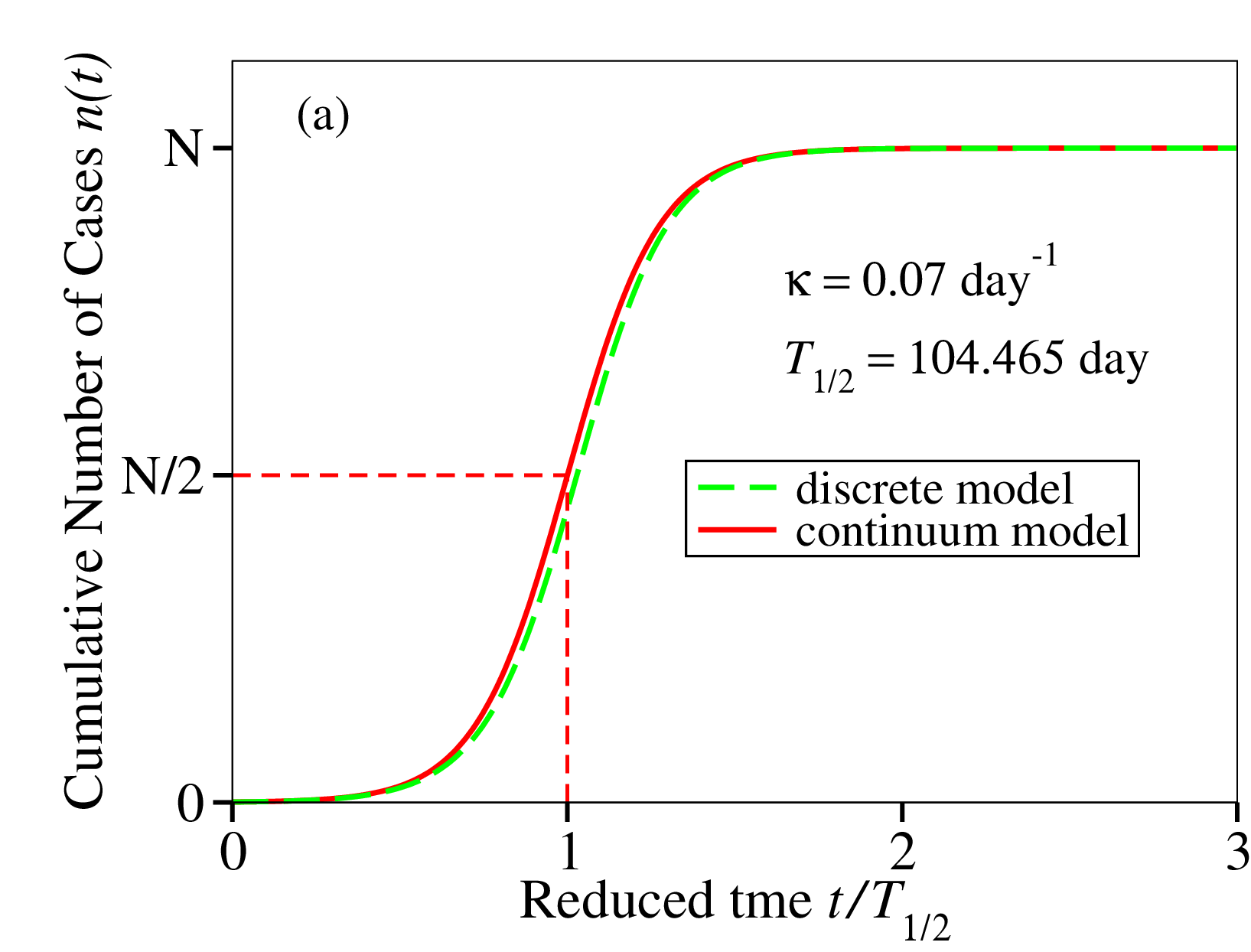}
    \hspace*{4ex}
    \includegraphics[width=0.45\textwidth]{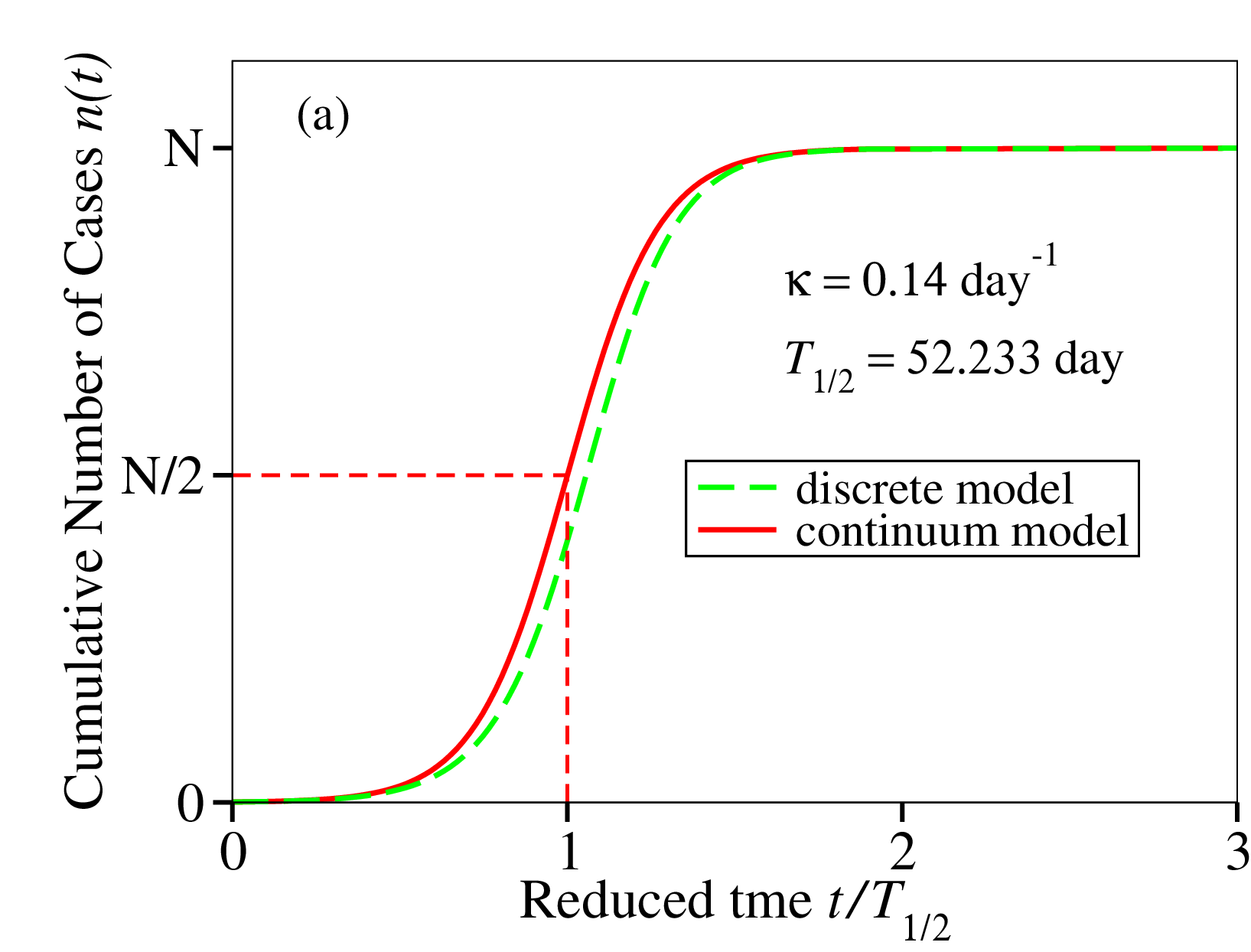}
  }
  $ $\\[5ex]
  \centerline{
    \includegraphics[width=0.45\textwidth]{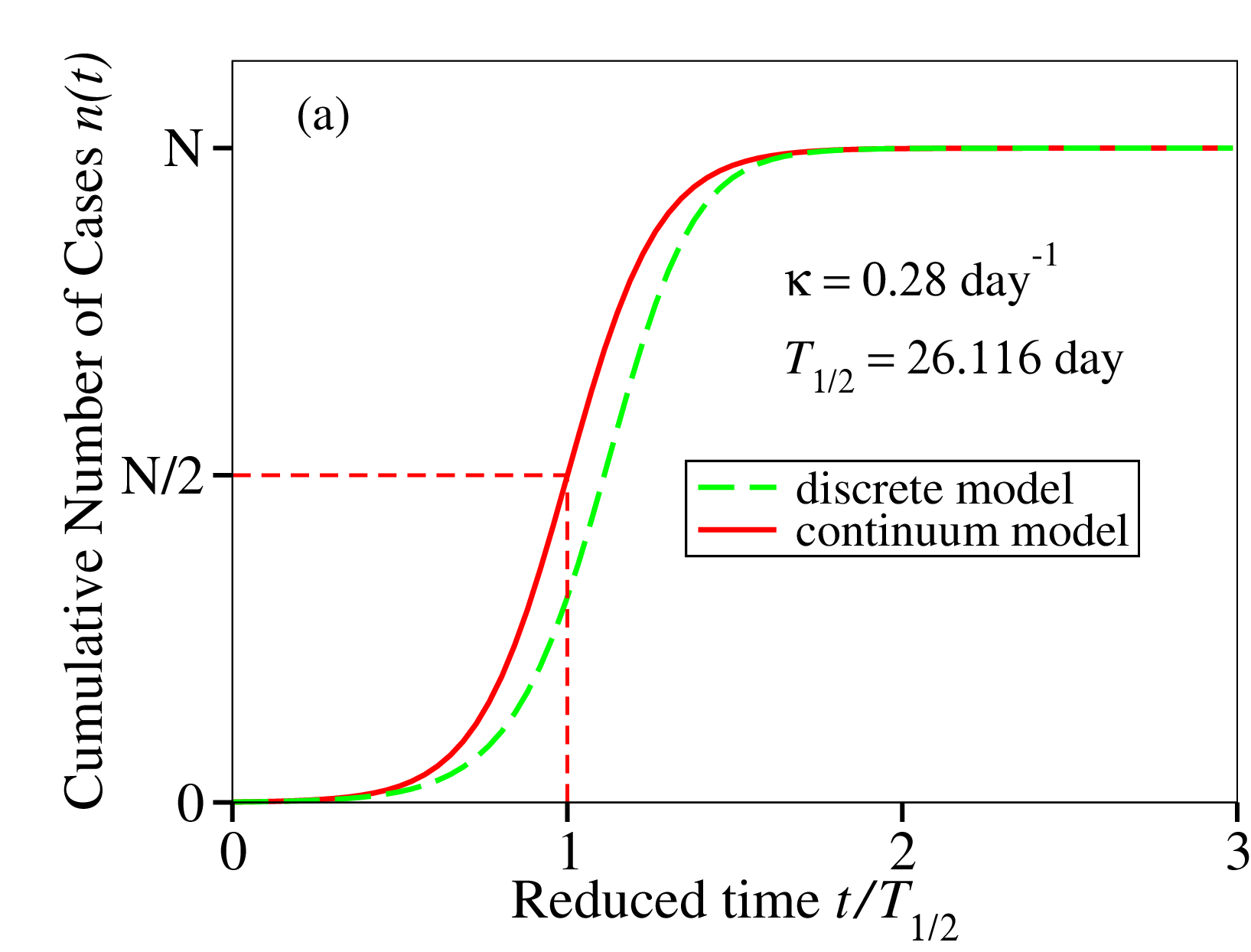}
    \hspace*{4ex}
    \includegraphics[width=0.45\textwidth]{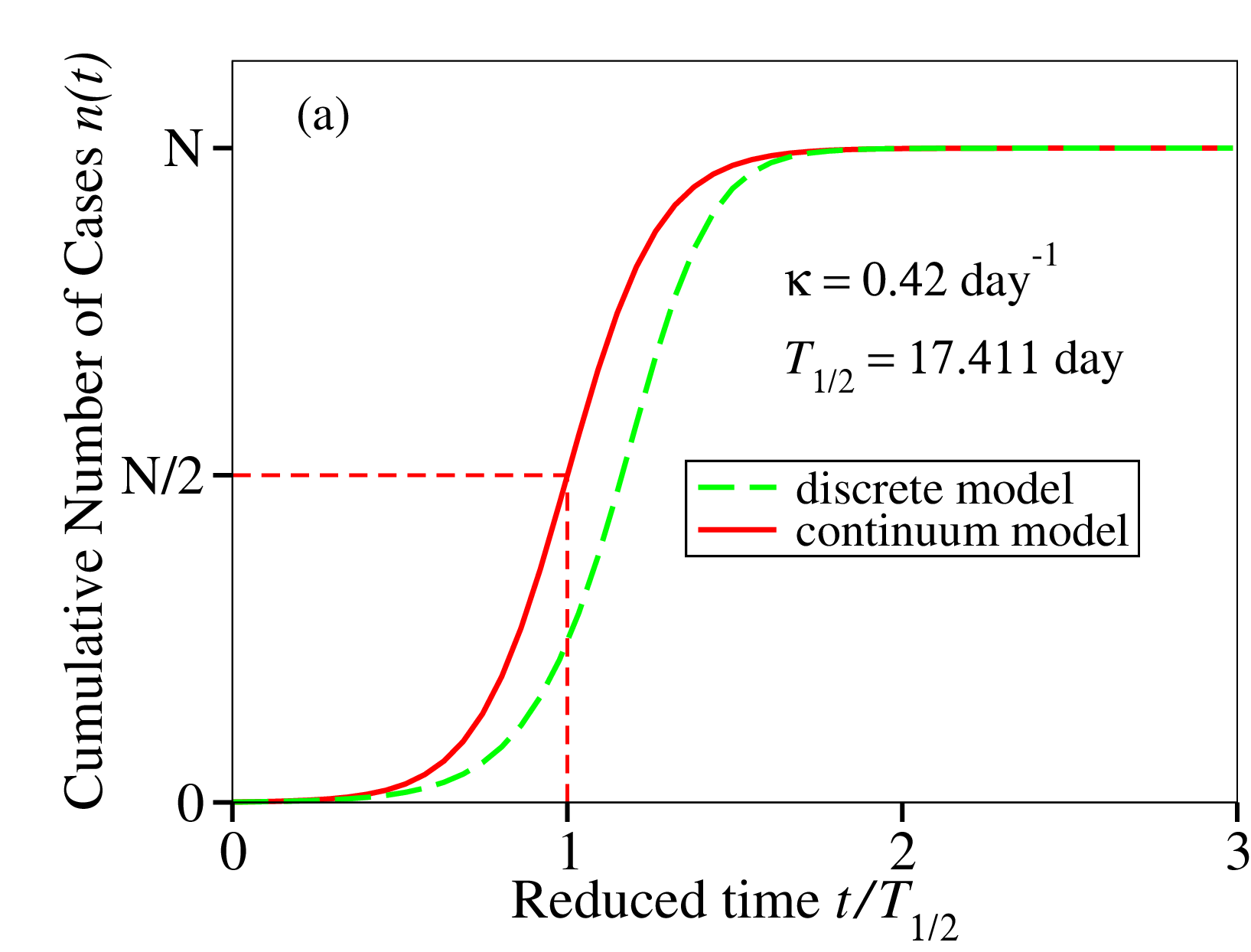}
  }
  \caption{Cumulative number of infection cases described by means of logistic growth models
    employing continuous and (``exact'') discrete time representation
    (\gls~(\ref{eq-n}) and (\ref{eq-dn-discrete}), respectively) for several typical values of
    the infection rate $\kappa$ indicated in the inset.
    Time on the $x$-axis is expressed in units of the half-time $T_{1/2}$.
  }
    \label{fig:f}
\end{figure}

\begin{figure}
  \centerline{
    \includegraphics[width=0.45\textwidth]{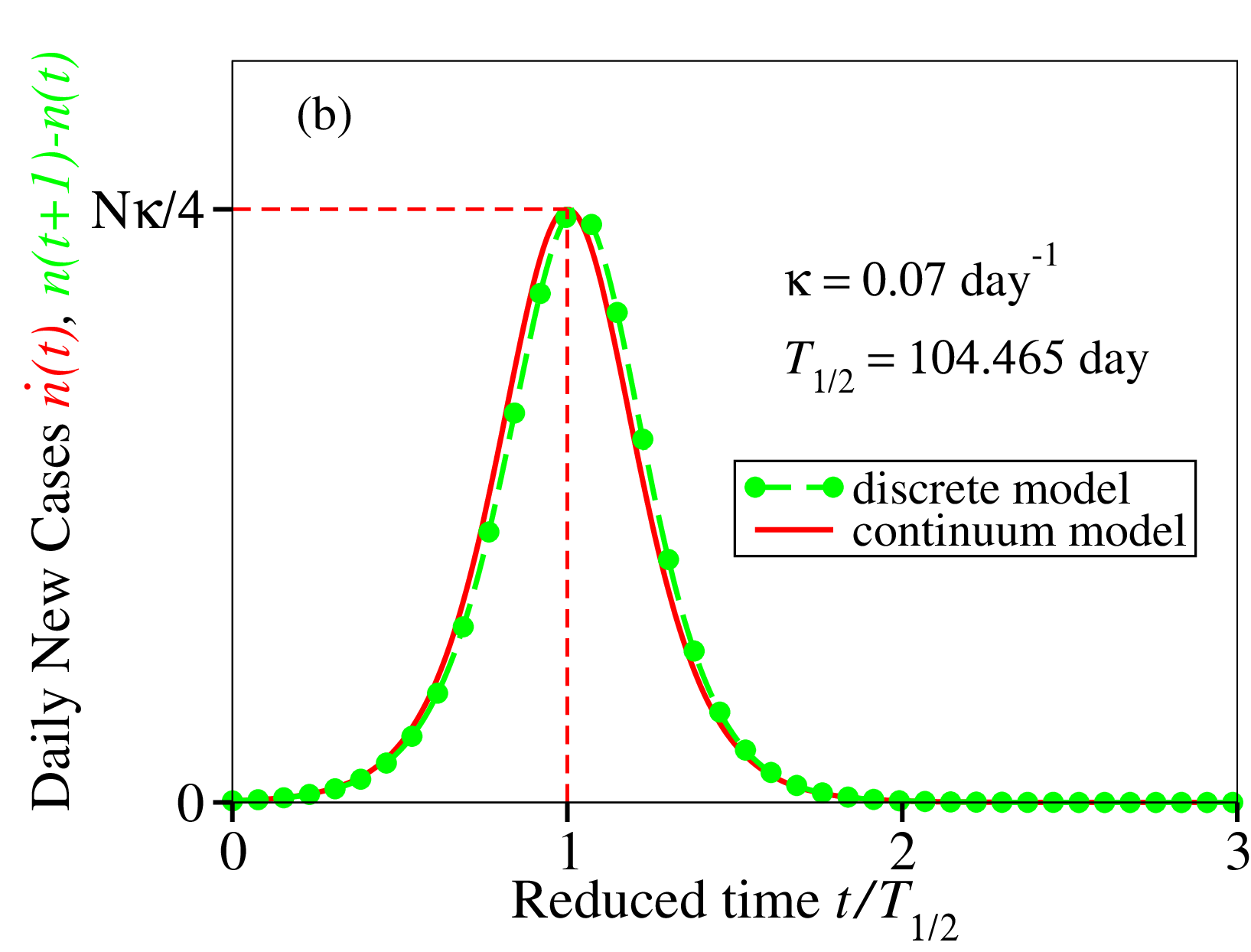}
    \hspace*{4ex}
    \includegraphics[width=0.45\textwidth]{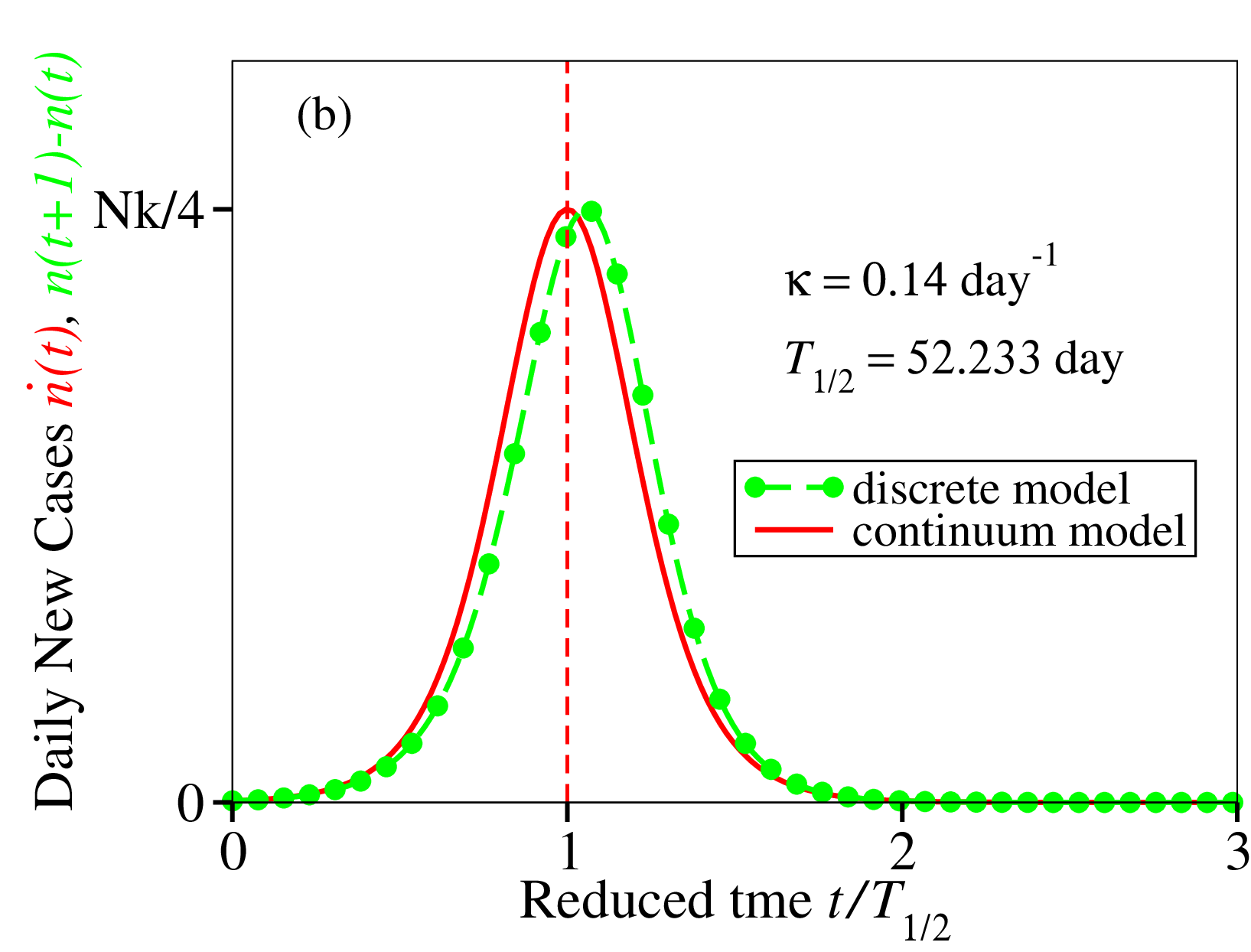}
  }
$ $\\[5ex]
  \centerline{
    \includegraphics[width=0.45\textwidth]{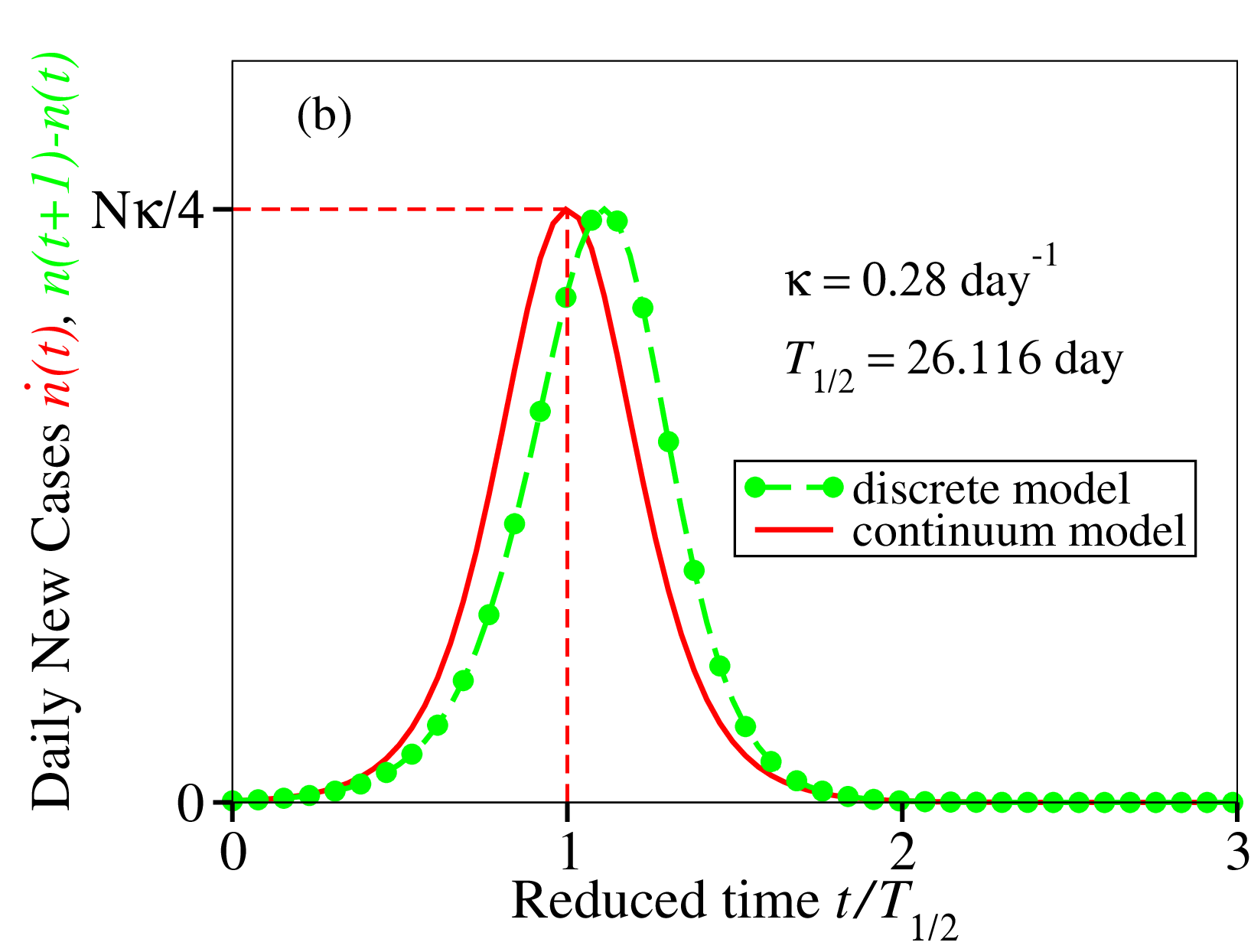}
    \hspace*{4ex}
    \includegraphics[width=0.45\textwidth]{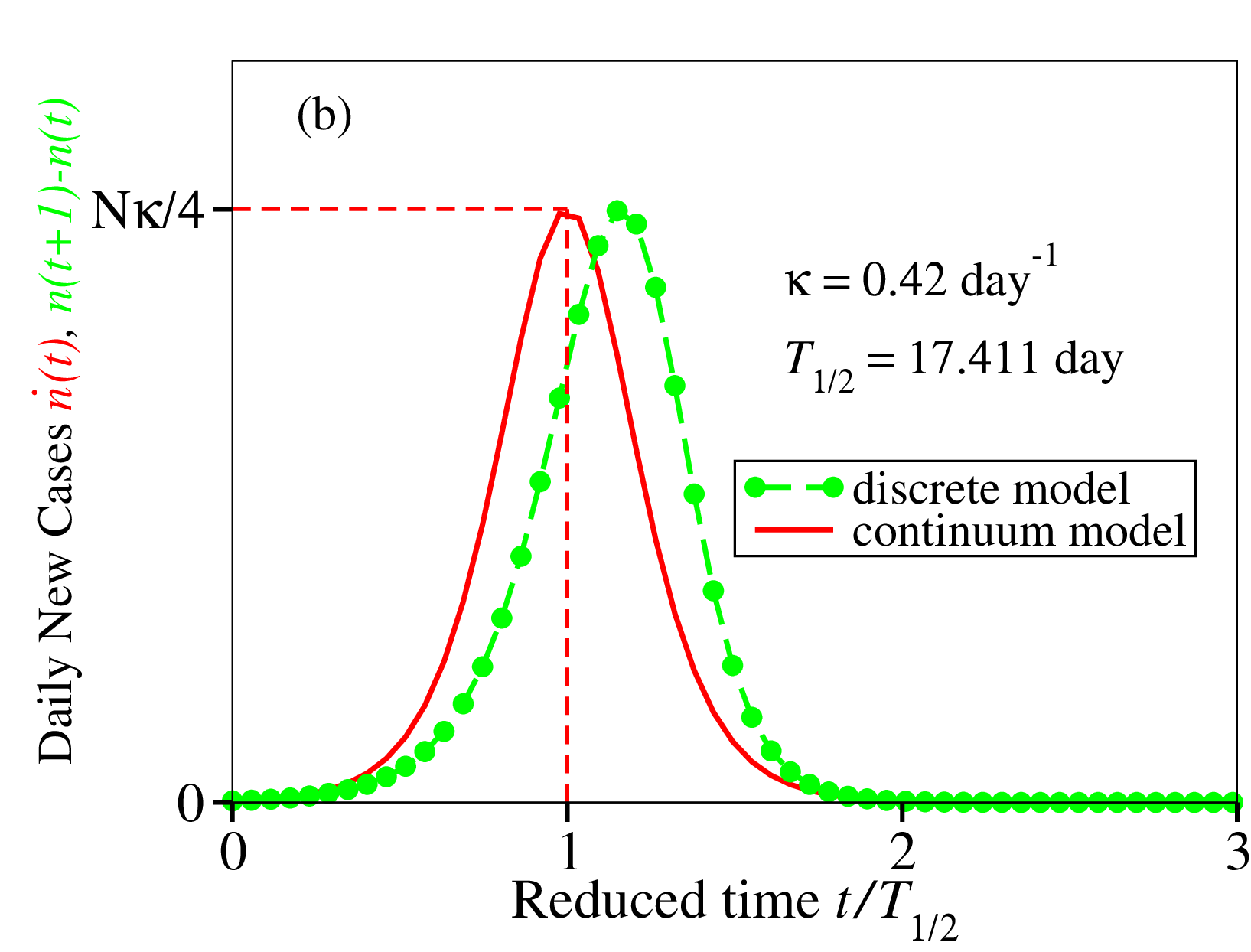}
  }
  \caption{Epidemiological curves depicting daily numbers of new infections
    described by means of logistic growth models
    employing continuous and (``exact'') discrete time representation
    (\gls~(\ref{eq-dn}) and \ref{eq-dn-discrete}), respectively)
    for several typical values of
    the infection rate $\kappa$ indicated in the inset.
    Time on the $x$-axis is expressed in units of the half-time $T_{1/2}$.
  }
    \label{fig:df}
\end{figure}

To start our analysis, we chose a value of infection rate
$\kappa = 0.14\,\mbox{day}^{-1}$, which turned out to describe a
regime of moderate restrictions during Slovenia COVID-19 epidemic
\cite{Baldea:2020f}. Results for the cumulative and daily
number of cases obtained within the continuous time and discrete time
using the value $\kappa = 0.14\,\mbox{day}^{-1}$ can be compared in \figsname\ref{fig:f}b and 
\ref{fig:df}b, respectively. These figures reveal a certain backward shift of
the continuous time curves with respect to the discrete time curves.

Based on naive intuition it seems reasonably to expect that situations corresponding
slower temporal variations (smaller infection rates $\kappa$)
can more appropriately be described quantitatively within
a continuous time representation than those characterized by faster variations
(larger $\kappa$ values). To check whether this is the case or not, we monitored
how results change by increasing $\kappa$. Results for the larger value $\kappa = 0.28\,\mbox{day}^{-1}$
are depicted in \figsname\ref{fig:f}c and \ref{fig:df}c.
They show backward shifts larger than for $\kappa = 0.14\,\mbox{day}^{-1}$. 
Further increasing $\kappa$ to the value 
$\kappa = 0.42\,\mbox{day}^{-1}$ (\figsname\ref{fig:f}d and \ref{fig:df}d) confirms this trend.
On the contrary, the curves computed for the smaller value
$\kappa = 0.07\,\mbox{day}^{-1}$ (\figsname\ref{fig:f}a and \ref{fig:df}a) 
reveal that the backward shift is smaller. So far, so good;
the intuitive expectation is supported by these calculations.
The curves based on continuous time exhibits a systematic shift backwards in time
whose magnitude varies monotonically with $\kappa$.
Noteworthily, the behavior displayed in \figsname\ref{fig:f} and \ref{fig:df}
refers to quantities plotted against the reduced time $t/T_{1/2}$.

\begin{figure}
  \centerline{\includegraphics[width=0.45\textwidth]{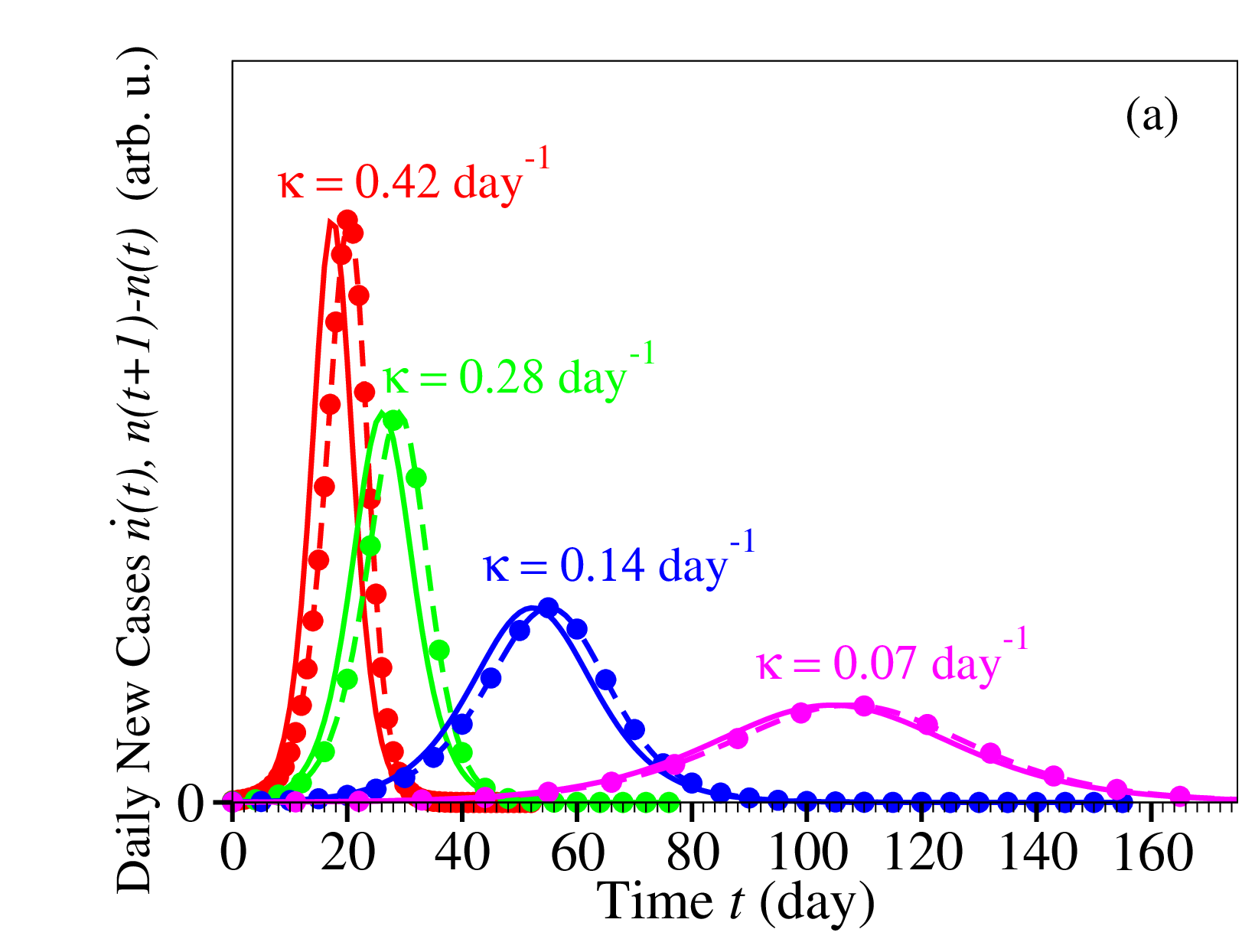}}
$ $\\[10ex]
  \centerline{\includegraphics[width=0.45\textwidth]{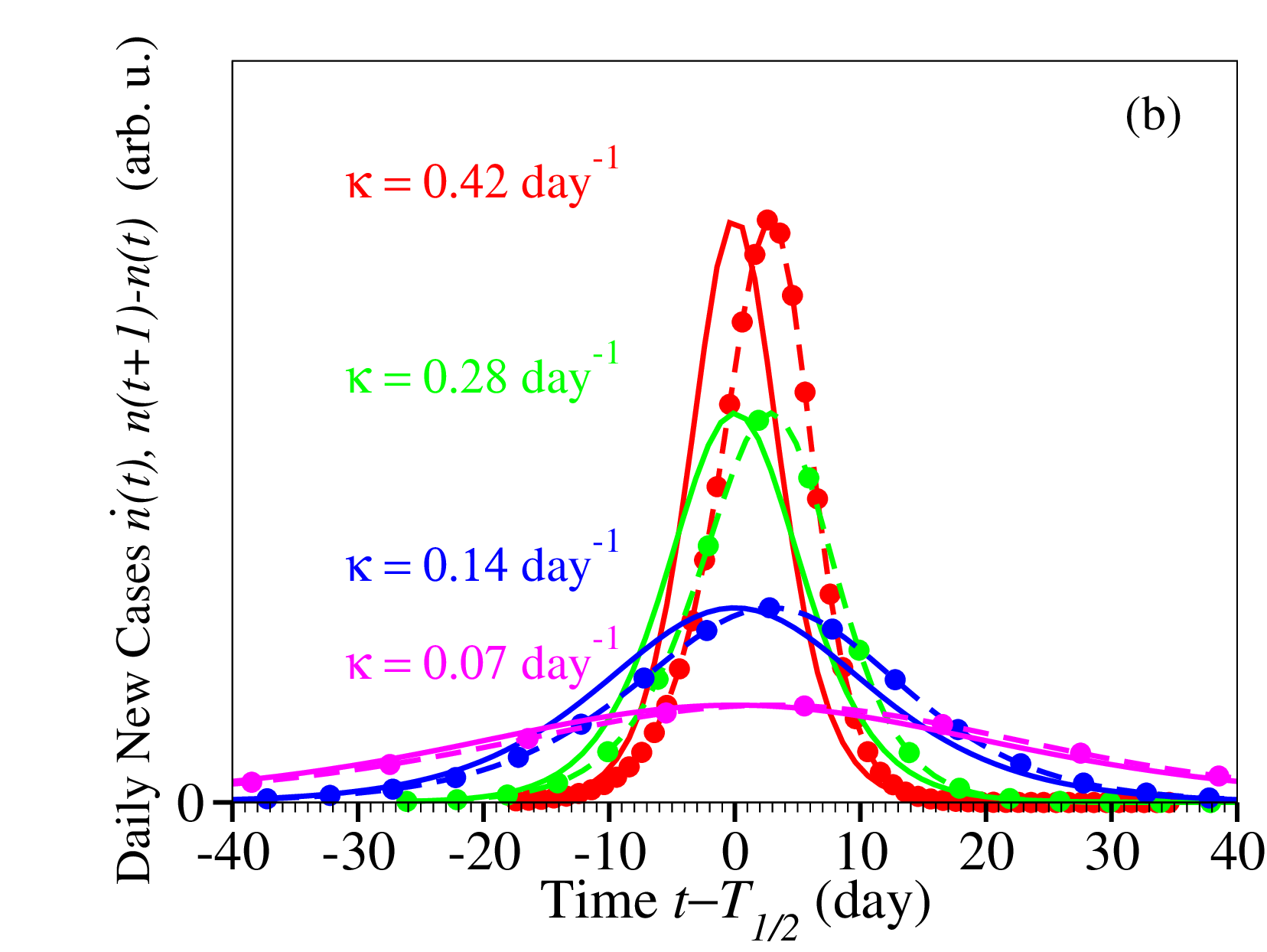}}
$ $\\[10ex]
  \centerline{\includegraphics[width=0.45\textwidth]{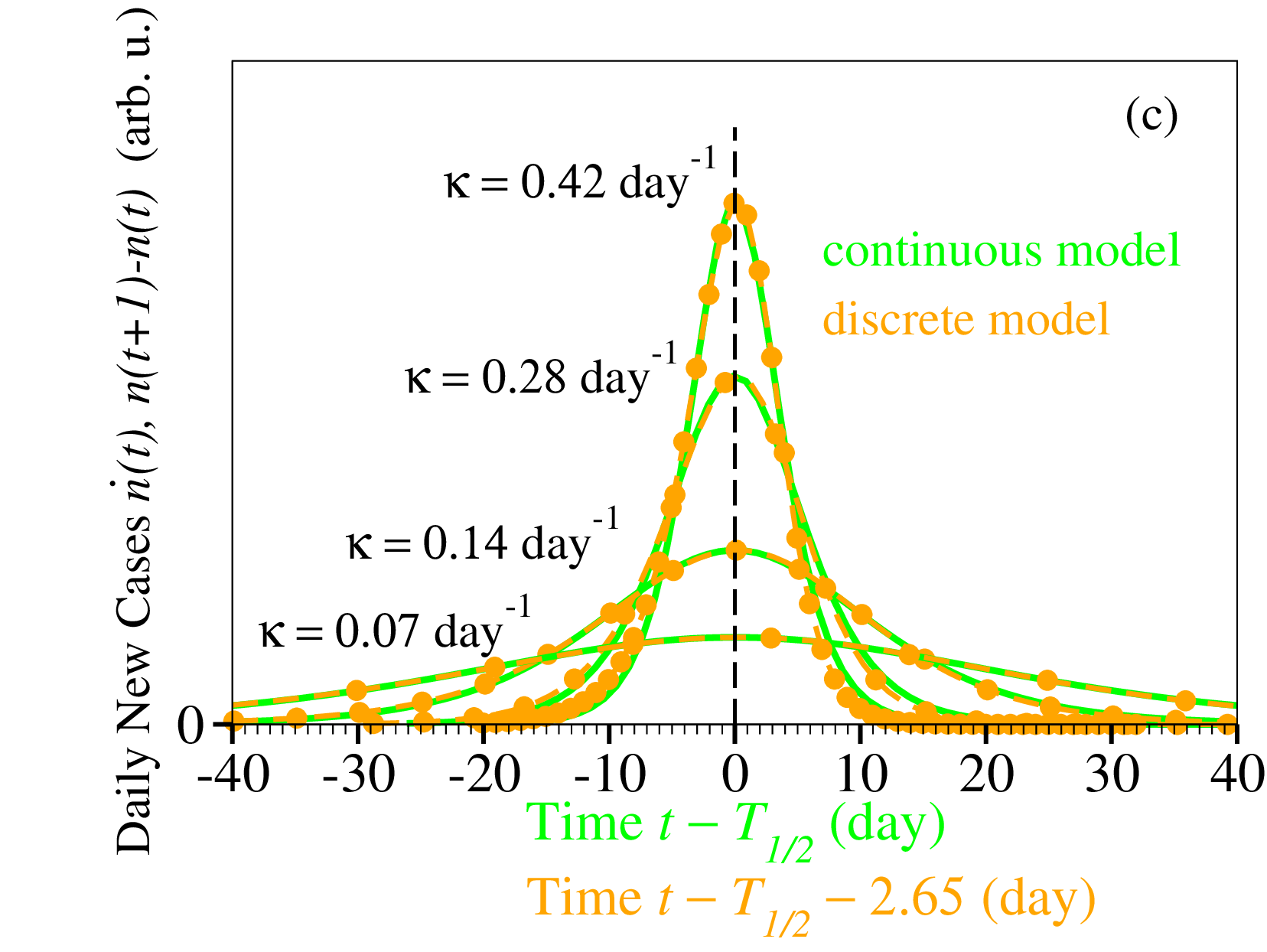}}
  \caption{Epidemiological curve depicting daily numbers of new infections
    described by means of logistic growth models
    employing continuous and (``exact'') discrete time representation for several typical values of
    the infection rate $\kappa$ indicated in the inset.
  }
    \label{fig:various-df}
\end{figure}

Surprise arises
when examining the curves plotted with the absolute time $t$ on the abscissa rather than the
reduced time  $t/T_{1/2}$.
The difference visible in \figsname\ref{fig:f} and \ref{fig:df}
is not so clear when using $t$ instead of $t/T_{1/2}$ on the x axis (see \figname\ref{fig:various-df}a). 
However, it becomes observable when the curves are shifted by $T_{1/2}$, that is,
are plotted versus $t - T_{1/2}$ (see \figname\ref{fig:various-df}b)
Still, the $\kappa$-dependence of the peak heights (they are proportional to $\kappa$, see
\gl~(\ref{eq-dn})) makes observation somewhat difficult.

Finally, the point we want to make is obvious in \figname\ref{fig:various-df}c.
For a very broad range of $\kappa$-values, the maximum of the
epidemiological curves deduced within the continuous time description 
is shifted backward in time by the $\kappa$-independent amount
$\tau \simeq -2.65$\,day while its height (=maximum number of daily new cases)
cannot be distinguished from that predicted within the (realistic) discrete time description. 
Concerning the $\kappa$-independence of $\tau$, one should note that
that the above value $\kappa = 0.42\,\mbox{day}^{-1}$ corresponds to a regime wherein infections were found
to growth exponentially in time while $\kappa = 0.07\,\mbox{day}^{-1}$
is almost half the $\kappa$-value at complete lockdown \cite{Baldea:2020f}.

To conclude, the above finding contradicts the expectation based on naive intuition
that faster population dynamics is poor(er)
describable in continuous time, or, put differently, that slower infections are better described within
a continuous time approach.

\section{A Technical Remark}
\label{sec-remark}
In arriving at \gl~(\ref{eq-dn-discrete}), we chose the forward (finite) difference $n(t+1)-n(t)$
to establish the correspondence between the discrete time and continuous
time versions of logistic growth. This procedure is just in the vein underlying the logistic growth
philosophy. The LHS of \gl~(\ref{eq-dn-continuous}) is nothing but the number of new daily cases. 
Via \gl~(\ref{eq-dn-continuous}), this number is assumed to be proportional to the number of total
of \emph{existing} infections $n(t)$ as long as the latter are few. Otherwise, 
the gradual increase in the fraction of the \emph{already}
infected individuals $n(t)/N$ reduces the probability for further infection. 

Purely mathematically speaking, as alternatives to the forward difference
($ d n(t)/ dt \to n(t+1) - n(t) $),
we could have chosen the backward finite difference ($ d n(t)/ dt \to n(t) - n(t-1) $)
or the symmetric finite difference ($ d n(t)/ dt \to \left[n(t+1) - n(t-1)\right]/2 $).

The backward difference leads to
\begin{equation}
\label{eq-dn-bd}
  n(t) - n(t-1) = \kappa n(t) \left[1 - \frac{n(t)}{N}\right]
\end{equation}
Although difference in numerical results obtained by employing \gl~(\ref{eq-dn-bd}) and 
\gl~(\ref{eq-dn-discrete}) are not notable, \gl~(\ref{eq-dn-bd}) does complain neither with the
idea of logistic growth (see above) nor to the causality principle.

Leading to 
\begin{equation}
\label{eq-dn-sd}
  \frac{n(t+1) - n(t-1)}{2} = \kappa n(t) \left[1 - \frac{n(t)}{N}\right]
\end{equation}
the symmetric finite difference (which is a better choice for
calculating numerical derivatives in other cases) is even more problematic here. In addition to
the  drawback noted in connection with \gl~(\ref{eq-dn-bd}), unrealistically,
the symmetric difference version
of \gl~(\ref{eq-dn-sd}) does not allow to uniquely predict time evolution $n(t)$ merely based
on a single initial condition ($n(0)=n_0$); \gl~(\ref{eq-dn-sd}) is a relationship between
populations at three consecutive days: $n(t+1)$, $n(t)$, and $n(t-1)$.
\section{Conclusion}
\label{sec:conclusion}
Based on the present results obtained within a logistic growth model, 
we conclude that the most notable inaccuracy brought about by
the continuous time description of epidemics --- which is mathematically more convenient than the
discrete time description --- is the prediction that the maximum number $\dot{n}$
of daily infections occurs earlier in time:  $T_{1/2}^{d} \to T_{1/2}^{c} < T_{1/2}^{d}$.
This backward shift in time $\tau \equiv T_{1/2}^{c} - T_{1/2}^{d} < 0$ (which does not affect
the height of the epi peak) amounts to
$\tau \approx -2.65$\,day, a robust estimate, which was found to be independent of the infection rate $\kappa$ 
ranging from large values deduced for situations
where COVID-19 infections growth exponentially in time 
to small values obtained at complete lockdown \cite{Baldea:2020f}.
This finding contradicts naive intuition expecting that faster infections (larger $\kappa$)
are poorer described within continuous time approaches than slower infections (smaller $\kappa$).

To end, let us emphasize that, although conducted in conjunction with the actual unprecedented
crisis caused by COVID-19 pandemic, the present study reports 
  results applying to logistic growth in general. No specific reference to a special
  real system was needed in the mathematical derivation presented above.
 
\label{lastpage}
\providecommand{\latin}[1]{#1}
\makeatletter
\providecommand{\doi}
  {\begingroup\let\do\@makeother\dospecials
  \catcode`\{=1 \catcode`\}=2 \doi@aux}
\providecommand{\doi@aux}[1]{\endgroup\texttt{#1}}
\makeatother
\providecommand*\mcitethebibliography{\thebibliography}
\csname @ifundefined\endcsname{endmcitethebibliography}
  {\let\endmcitethebibliography\endthebibliography}{}

\end{document}